\documentclass[%
reprint,
superscriptaddress,
amsmath,amssymb,
aps,
pre,
]{revtex4-1}
\usepackage{graphicx}
\usepackage{float}
\usepackage[caption=false]{subfig}
\usepackage{dcolumn}% Align table columns on decimal point
\usepackage{bm}
\newcommand{\mathsym}[1]{{}}
\newcommand{\unicode}[1]{{}}

\begin{document}

\title{Why large icosahedral viruses need scaffolding proteins: The interplay of Gaussian curvature and disclination interactions.}

\author{Siyu Li}
\affiliation{Department of Physics and Astronomy,
University of California, Riverside, California 92521, USA}
\author{Polly Roy} 
\affiliation{Department of Pathogen Molecular Biology, Faculty of Infectious and Tropical Diseases, London School of Hygiene and Tropical Medicine, London WC1E 7HT, UK}
\author{Alex Travesset}
\affiliation{Department of Physics and Astronomy, Iowa State and Ames Lab, Ames, Iowa 50011-3160, USA}
\author{Roya Zandi}
\affiliation{Department of Physics and Astronomy,
University of California, Riverside, California 92521, USA}

% \significancestatement{Despite a plethora of experimental data on the role of scaffolding proteins in the structure of viral shells, no theoretical/numerical explanations have been put forward to decipher the indisputable need of large shells for scaffolding proteins. The paper explains the underlying physical mechanisms for the formation of large viral shells and elucidates the ``universal'' role of scaffolding proteins in the formation of large spherical crystals.}
% \authorcontributions{A.T., P.R. and R.Z. designed the research. A.T., S.L. and R.Z. developed the models, performed the analytical calculations and designed the simulations. S.L. performed the simulations. R.Z., A.T. and S.L. analyzed the data and interpreted the results. A.T., R.Z. and S.L. wrote the paper with support from P.R..}
% \authordeclaration{The authors declare no conflict of interest.}
% \correspondingauthor{\textsuperscript{1}To whom correspondence should be addressed. E-mail: sli032@ucr.edu}
% \keywords{self-assembly $|$ scaffolding proteins$|$ continuum elasticity theory} 

\begin{abstract}
While small single stranded viral shells encapsidate their genome spontaneously, many large viruses, such as the Herpes virus or Infectious Bursal Disease Virus (IBDV), typically require a template, consisting of either scaffolding proteins or inner core. Despite the proliferation of large viruses in nature, the mechanisms by which hundreds or thousands of proteins assemble to form structures with icosahedral order (IO) is completely unknown. Using continuum elasticity theory, we study the growth of large viral shells (capsids) and show that a non-specific template not only selects the radius of the capsid, but leads to the error-free assembly of protein subunits into capsids with universal IO. We prove that as a spherical cap grows, there is a deep potential well at the locations of disclinations that later in the assembly process will become the vertices of an icosahedron. Furthermore, we introduce a minimal model and simulate the assembly of viral shell around a template under non-equilibrium conditions and find a perfect match between the results of continuum elasticity theory and the numerical simulations. Besides explaining available experimental results, we provide a number of new predictions. Implications for other problems in spherical crystals are also discussed.
\end{abstract}

\maketitle

\begin{figure*}
 \centering
  \includegraphics[width=\linewidth]{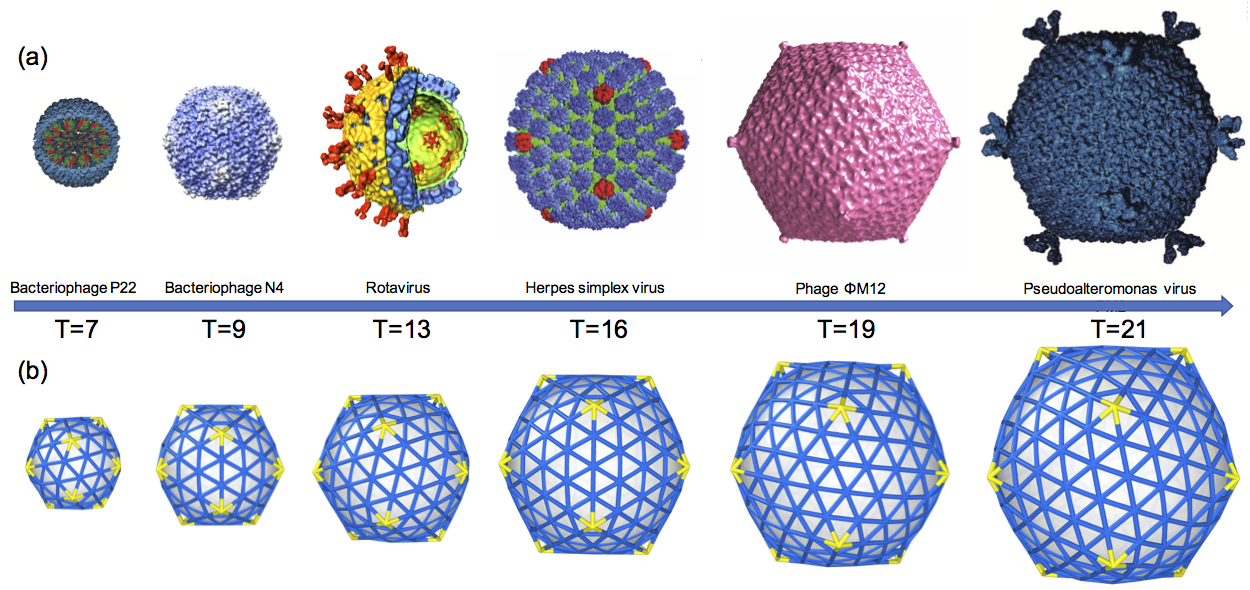}
  \caption{\footnotesize 
Figure 1a. From left to right: Bacteriophage P22~\cite{jordan2016}, Bacteriophage N4~\cite{choi2008}, Rotavirus~\cite{pesavento2003}, Herpes simplex virus~\cite{zhou2000}, Phage $\Phi$M12~\cite{stroupe2014} and Pseudoalteromonas virus~\cite{oksanen2017}. The triangulation number of each virus is shown below it. The scaffolding proteins and hydrogenases inside the capsid of Bacteriophage P22 and the inner shell of Rotavirus are illustrated in the figure. To form structures with IO, all viruses in the figure need scaffolding proteins as illustrated for Bacteriophage P22. Only Rotavirus requires a preformed scaffolding layer. Rotavirus belong to Reoviridae virus family, they all form $T=13$ and have multi-shell structures.
Figure 1b. Capsids obtained in the simulations from left to right: $T=7$, $T=9$, $T=13$, $T=16$, $T=19$ and $T=21$.   }
\label{gallery}
\end{figure*}

More than fifty years ago, Caspar and Klug~\cite{CASPAR1962} made the striking observation that the capsids of most spherical viruses display icosahedral order(IO), defined by twelve five coordinated units (disclinations or pentamers) occupying the vertices of an icosahedron surrounded by hexameric units, see Fig.~\ref{gallery}. While many studies have shown that this universal IO is favored under mechanical equilibrium~\cite{bruinsma,Fejer:10,Rapaport:04a}, the mechanism by which these shells grow, circumventing many possible activation barriers, and leading to the perfect IO remains mainly unknown.

Under many circumstances, small icosahedral capsids assemble spontaneously around their genetic material, often a single-stranded viral RNA~\cite{elife,Comas,Cornelissen2007,Sun2007,Nature2016}. Yet, larger double-stranded (ds) RNA or DNA viruses require what we generically denote as the template: scaffolding proteins (SPs) or an inner core~\cite{Thuman-Commike1998,Earnshaw1978,Aksyuk2015,Saad1999,Saugar2010,Coulibaly2005}. The focus of this paper is on these large viruses that require a template for successful assembly.

The major difficulty in understanding the pathway towards IO is apparent from the results of the generalized Thomson problem, consisting of finding the minimum configuration for interacting $M$-point particles constrained to be on the surface of a sphere. Simulation studies show that the number of metastable states increase exponentially with $M$~\cite{Erber1991}, and only with the help of sophisticated optimization algorithms at relatively small values of $M$~\cite{Morris1996, Zandi2004,elife,Zandi2016}, it is possible to obtain IO ground states. These situations, typical of spherical crystals, become even more difficult when considering the assembly of large capsids, in which once protein subunits are attached and a few bonds are made, it becomes energetically impossible for them to re-arrange: Should a single pentamer appear in an incorrect location, IO assembly would fail.  

The combined effect of irreversibility and the inherent exponentially large number of metastable states typical of curved crystals puts many drastic constraints on IO growth. The complexity of the problem may be visualized by the various viral shells illustrated in Fig.~\ref{gallery}, characterized by a structural index, the T number~\cite{CASPAR1962,Wagner2015,Chen:2007b,Vernizzi:2007a} $T=h^2+k^2+hk$, 
% \begin{equation}\label{Eq:T_Num}
% T=h^2+k^2+hk
% \end{equation}
with $h$ and $k$ arbitrary integers, such that the crystal includes $60T$ monomers or $10(T-1)$ hexamers and 12 pentamers (disclinations).

A possible mechanism to successfully self-assemble a desirable structure might consist of protein subunits with chemical specificity, very much like in DNA origami~\cite{Rothemund2006} where structures with complex symmetries are routinely assembled. In viruses, however, capsids are built either from one or a few different types of proteins, so specificity cannot be the driving mechanism leading to IO~\cite{Yu2013,Nature2016,nguyen2006continuum,Chen:2007b,Zandi2004}.  In this paper, we show that a ``generic'' template provides a robust path to self- assembly of large shells with IO. This is consistent with many experimental data in that regardless of amino acid sequences and folding structures of virus coat and/or scaffolding proteins, due to the ``universal'' topological and geometrical constraints, large spherical viruses need scaffolding proteins to adopt IO, see Fig.~\ref{gallery}.  Although the focus of the paper is on virus assembly, the implication of our study goes far beyond and extend to many other problems where curved crystals are involved, a point that we we further elaborate in the conclusion~\cite{Lidmar2003,vernizzi2011platonic}.

The distinct feature of spherical crystals is that their global structure is constrained by topology. More concretely, if $s({\bf x})$ is the disclination density, then 
\begin{equation}\label{Eq:Euler}
\int d^2{\bf x} ~s({\bf x}) =  2\pi \chi \ ,
\end{equation}
where $\chi$ is the Euler characteristic ($\chi=2$ for a sphere). However a capsid closes only at the end of the assembly, and thus,
Eq.~\ref{Eq:Euler} does not really restrict the number of disclinations during the growth process, as pentamers or other disclinations may be created or destroyed at the boundaries. For a complete shell, the easiest way to fulfill Eq.~\ref{Eq:Euler} is with twelve $q=+\frac{\pi}{3}$ disclinations, and this is the case we will follow hereon. 
%The possibility of satisfying Eq.~\ref{Eq:Euler} with additional disclinations has been discussed elsewhere\cite{Bowick2000}, and as we clarify below, it is not relevant to this study.

%\begin{figure}
% \centering
%\subfloat[]{\includegraphics[width=0.35\linewidth]{t13.png}} 
%\subfloat[]{\includegraphics[width=0.45\linewidth]{rota.png}}
%\subfloat[]{\includegraphics[width=0.45\linewidth]{p22.png}}
%\caption{\footnotesize T=13 shell}
%\label{shellenergy}
%\end{figure}

% EMDB EMD-1509
% Figure from Eric Mossel , Mary Estes and Frank Ramig
% EMDB EMD-5718

A minimal model for spherical crystals consists of a free energy
\begin{eqnarray}\label{Eq:Free energy}
F_c &=& \int d^2{\bf x} \left[ \mu u_{\alpha \beta}^2 + \frac{\lambda}{2} (u_{\alpha \alpha})^2 \right]\ + \frac{\kappa}{2}\int d^2{\bf x} (H({\bf x})- H_0)^2
\nonumber\\
    &\equiv& F_c^l  + F_c^b 
\end{eqnarray}
where $u_{\alpha \beta}$ is the strain tensor. The coefficients $\mu, \lambda$ are the Lame coefficients, which depend on the microscopic underlying interactions. Here $H({\bf x})$ is the extrinsic curvature of the template, $H_0$ the spontaneous curvature and $\kappa$ the bending rigidity. By integrating the phonon degrees of freedom, we can recast the term $F^{l}_c$ in Eq.~\ref{Eq:Free energy} as a non-local theory of interacting disclinations, with free energy~\cite{Bowick2000}
\begin{eqnarray}\label{Eq:Free energy_defects}
F_c^l &=& \frac{K_0}{2}\int d^2{\bf x} d^2{\bf y} \left[(K({\bf x})-s({\bf x}) )G({\bf x},{\bf y}) \right. \times \nonumber\\
  &\times& \left. (K({\bf y})-s({\bf y})) \right]\ ,
\end{eqnarray}
where $K({\bf x})$ is the Gaussian curvature and $K_0$ is the Young modulus. The disclination density $s({\bf x}) = \sum_{i=1}^{12}q_i \delta({\bf x}- {\bf x}_i)$ has as variables the positions of 12 disclinations, each of charge $q_i = \frac{\pi}{3}$. The function $ G({\bf x},{\bf y})$ is the inverse of the Laplacian square~\cite{Bowick2000}. All previous studies for the model in Eq.~\ref{Eq:Free energy_defects} have been done for curved crystals without a boundary. In this paper, we provide, for the first time, the necessary formalism to include the presence of a boundary.

A discrete version of Eq.~\ref{Eq:Free energy} is given by~\cite{seung1988defects,vernizzi2011platonic,Bowick2000,Lidmar2003}
\begin{equation}\label{Eq:discrete}
F_d = E_s + E_b = \sum_{i} \frac{1}{2}k_s(b_{i}-b_0)^2 + \sum_{i,j}k_b[1-\cos(\theta_{ij}-\theta_0)]
\end{equation}
with $\theta_0$ a preferred angle, related to the spontaneous curvature $H_0$. The stretching energy sums over all bonds $i$ with $b_0$ the equilibrium bond length and the bending energy is between all neighboring trimers indexed with $ij$. We further assume that there is an attractive force between the trimers and the preformed scaffolding layer (inner core) (see Fig.\ref{largeT}), which, consistent with our minimal model, involves a simple LJ-potential $E_{LJ}=\sum_{i}4 \epsilon [(\frac{\sigma}{r_{i}})^{12}-2(\frac{\sigma}{r_{i}})^6 ]$ with $\epsilon$ the depth of the potential and $\sigma$ the position of minimum energy corresponding to optimal distance between the center of the core and subunits. In the next section, we associate a dynamics to these models, which corresponds to following a local minimum energy pathway. 

\section*{Methods}\label{methods}

\subsection*{Discrete model}\label{discrete}

The growth of the shells is based on the following assumptions~\cite{Comas,Nature2016,Zandi2016,Yu2013}: At each step of growth, a new trimer is added to the location in the boundary which makes the maximum number of bonds with the neighboring subunits. This is consistent with the fact that protein-protein attractive interaction is weak and a subunit can associate and dissociate till it sits in a position that forms a few bounds with neighboring proteins. These interactions eventually become strong for the subunits to dissociate and trimer attachment becomes irreversible~\cite{elife}. The attractive interactions between subunits, whose strength depends on electrostatic and hydrophobic forces, are implicit in the model. Note that pH and salt can modify the strength of protein-protein and protein-template interactions and thus the growth pathway. The impact of pH and salt on the shell assembly will be pursed elsewhere. 

\begin{figure}[H]
\includegraphics[width=\linewidth]{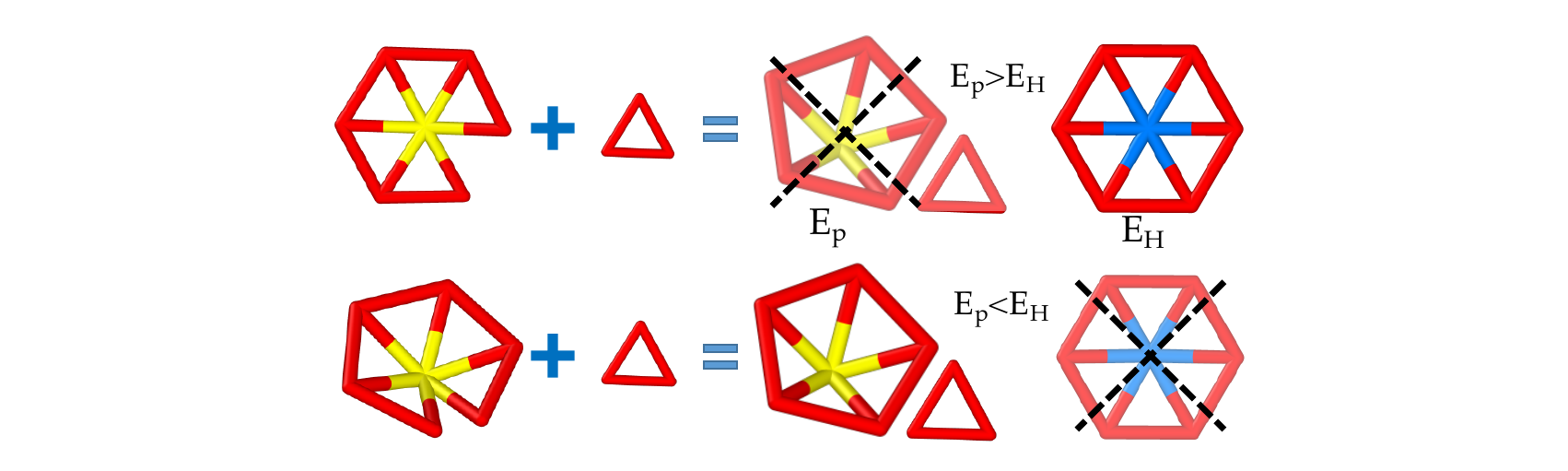}
  \caption{\footnotesize 
  Dynamics of formation of a hexamer vs. a pentamer: five trimers are attached at a vertex with an opening angle close to $\pi /3$ at the top and much smaller than  $\pi /3$ at the bottom. If the energy per subunit of formation of a pentamer $E_p$ is higher than a hexamer $E_H$, then a hexamer forms (top); otherwise, a pentamer assembles (bottom).
  }
  \label{PHchoice}
\end{figure}
A crucial step in the assembly process is the formation of pentamers, which occurs only if the local energy is lowered, as illustrated in Fig.~\ref{PHchoice}. After the addition of each subunit or the formation of a pentamer, using HOOMD package~\cite{hoomd1,hoomd2}, we allow the triangular lattice to relax and to find its minimum energy configuration~\cite{Wagner2015}. 

The proposed mechanism follows a sequential pathway where trimers ($T$) attach to the growing capsid ($C\rightarrow C^{\prime}$) according to the reaction
\begin{eqnarray}\label{Eq:reaction}
T+C & \leftrightarrows& TC \nonumber \\
TC & \rightarrow &  C^{\prime} 
\end{eqnarray}
with characteristic rates $k_D, k^{\prime}_D$ and $k_r$. The rate $k_D=2\pi D_T R_T $ is diffusion limited, with $R_T$ the trimer radius and $D_T$ its diffusion coefficient, so that the reaction speed is linear in trimer concentration $v_{TC}=k_D [T]$, $k_D^{\prime}$ is the detachment rate as the trimer searches for the local minimum, and $k_r$ is the irreversible rate of attachment of the trimer to the capsid. The combined reaction rate is therefore $k_T = \frac{k_r k_D}{k^{\prime}_D + k_r}$. Once the second reaction in Eq.~\ref{Eq:reaction} takes place, there is no possibility for correcting mistakes: if a pentamer forms in the incorrect location, IO is frustrated. With some additional assumptions about the dependence of $k_r$ on the coordination of the growing capsid, it is possible to derive overall rates for capsid formation, a problem that will be pursued elsewhere. 

Two important parameters arises in discussing spherical crystals with the model Eq.~\ref{Eq:discrete}. One is the Foppl von-Karman (FvK) number~\cite{Lidmar2003}
\begin{equation}\label{Eq:FvK}
\gamma=\frac{b_0^2k_s }{k_b} \ ,
\end{equation}
which measures the ratio of stretching to bending moduli. When the FvK number is large, the protein subunits optimize stretching and bend away from their preferred radius of curvature showing some degree of faceting, which is the case of large viruses, see Figure~\ref{gallery}. For the case of template driven self-assembly, we introduce a new parameter 
\begin{equation}
\eta =  \frac{k_b}{\epsilon},
\end{equation}
which measures the relative strength of the bending rigidity to the attraction of the trimers to the template. For small $\eta$, the proteins follow the core curvature during growth at all the time, regardless of proteins spontaneous curvature. For large $\eta$, the shell detaches from the core and follow its preferred curvature. In this paper we will be mostly interested in the regime $\eta \approx {\cal O}(1)$ and $\gamma \gg 1$, where the template, rather than the spontaneous curvature dictates the size of the capsid.

\subsection*{Continuum model}\label{continuum}

We now consider the model given in Eq.~\ref{Eq:Free energy_defects} on a spherical cap with an aperture angle $\theta_m$, so that its geodesic radius is $R_m=\theta_m R$, see Fig.~\ref{largeT}b. The Lame term ($F_c^l$) in Eq.~\ref{Eq:Free energy_defects} can then be written as 
\begin{equation}\label{F1}
F_c^l=\frac{1}{2 K_0}\int d^2{\bf x} \sqrt{g}\left(\Delta\chi\right)^2 \ ,
\end{equation}
where $g_{\mu \nu}$ is the metric defining the surface and
the Laplacian is $\Delta = -\frac{1}{\sqrt{g}}\partial_{\mu} g^{\mu \nu}\partial_{\nu}$, with $\chi$ the Airy Stress function that satisfies
\begin{eqnarray}\label{Eq:bieq}
\frac{1}{K_0}\Delta ^2\chi({\bf x}) &=&  s({\bf x})- K({\bf x}) .
\end{eqnarray}
In SI Appendix, we provide the detailed calculations. We note that approximate solutions of Eq.~\ref{Eq:bieq} are available under the assumption that the Laplacian is computed with a flat metric, see Ref~\cite{Grason2012}, which immediately leads to
%\begin{equation}\label{Eq:EM_Topo}
$\int d^2 {\bf x} K({\bf x}) = \int \frac{d^2{\bf x}}{R^2} =  \frac{A}{R^2} =  \pi \neq 2\chi \pi = 4\pi,$
%\end{equation}
directly violating the topological constraint Eq.~\ref{Eq:Euler}. Therefore previous results~\cite{Castelnovo2017} are limited to small curvatures or aperture angles ($\theta_m \ll \pi$). The generalization of Eq.~\ref{Eq:Free energy_defects} to include boundaries proceeds by defining the stress tensor by the expression $\sigma^{\alpha \beta} = g^{\alpha \beta} {\Delta} \chi({\bf x}) - g^{\alpha\mu}g^{\beta\nu}  {\nabla}_{\mu}{\nabla}_{\nu} \chi({\bf x})$.
% \begin{equation}\label{Eq:form_linear_stress_ref}
% \sigma^{\alpha \beta} = g^{\alpha \beta} {\Delta} \chi({\bf x}) - g^{\alpha\mu}g^{\beta\nu}  {\nabla}_{\mu}{\nabla}_{\nu} \chi({\bf x}) \ .
% \end{equation}
We now include a stress free condition $\sigma_{\alpha \beta} n^{\beta}=0$ at the boundary, where $n^{\alpha}$ is the normal to the boundary. For a spherical cap, see Fig.~\ref{largeT}b, we use the metric $ds^2 = g_{\mu \nu} dx^{\mu} dx^{\nu}= dr^2 + R^2\sin^2(r/R) d\phi^2$.  Note that following the simulation outcomes, we ignore boundary fluctuations. This is mainly because of the strength of protein-protein interactions and line tension implicit in the growth model and is consistent with the simulation results.

% \begin{equation}\label{Eq:metric_sp_cap}
% ds^2 = g_{\mu \nu} dx^{\mu} dx^{\nu}= dr^2 + R^2\sin^2(r/R) d\phi^2 \ .
% \end{equation}

With the above definitions, the topological constraint Eq.~\ref{Eq:Euler} is satisfied exactly for a sphere. The free energy Eq.~\ref{Eq:Free energy_defects} then becomes
\begin{eqnarray}\label{Eq:Free energy_defects_simple}
F_c^l(\theta_m, {\bf x}_i) & = & E_0(\theta_m) + \sum_{i=1}^N E_{0d}({\bf x}_i, \theta_m) + \nonumber\\
&+& \sum_{i=1}^N \sum_{j=1}^N \hat{E}_{dd}({\bf x}_i,{\bf x}_j, \theta_m) 
\end{eqnarray}
with $E_0$ is the free energy of the hexamers, $E_{0d}$ the interplay between Gaussian curvature and pentamers and $\hat{E}_{dd}$ describes disclination(pentamer) interactions. It is convenient to separate this last term as
\begin{eqnarray}\label{Eq:dd_term}
F_c^{dd} &=& \sum_{i=1}^N \sum_{j=1}^N \hat{E}_{dd}({\bf x}_i,{\bf x}_j, \theta_m)
\nonumber\\
% &=&\sum_{i=1}^N E_{dd}({\bf x}_i,{\bf x}_i) + \sum_{i=1}^N \sum_{j>i}^N E_{dd}({\bf x}_i, {\bf x}_j) \nonumber \\
&=& \sum_{i=1}^N E_{self}({\bf x}_i) + \sum_{i=1}^N \sum_{j>i}^N E_{dd}({\bf x}_i, {\bf x}_j),
\end{eqnarray}
where $E_{self}({\bf x}_i)$ is the disclination self-energy, which depends on the location of a pentamer relative to the boundary.

\section*{Results}

\begin{figure*}
	\centering
	\includegraphics[width=\textwidth]{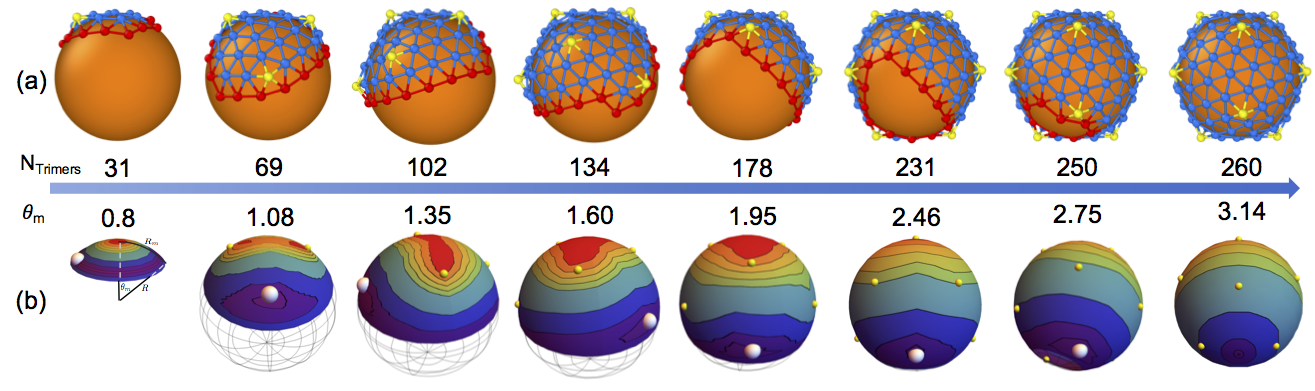}
	\caption{\footnotesize 
		The snapshots of a T=13 growth in discrete simulation (first row) and continuum theory (second row). The upper caps correspond to the simulation growth with triangles representing the trimers. The yellow vertices belong to pentamers, blue ones to hexamers and red ones to the cap edge. The gold core mimics the preformed scaffolding layer or inner core. The lower caps denote the energy contourplots for the newest disclinations that appear in the purple energy well, with geodesic shell size $R_m=R\theta_m$. The red region has the highest energy and purple the lowest one. There is a yellow ball in the position of each disclination. The largest ball corresponds to a newly formed disclination.}
	\label{largeT}
\end{figure*}

Consistent with the assumptions describing the dynamics of growth noted in previous section, we consider the spherical cap in Fig.~\ref{largeT}b with an aperture angle, which monotonically varies from $\theta_m=0$ to $\theta_m=\pi$ (sequential growth) as a function of time $\theta_m(t)$. Then, for each value of $\theta_m$ we calculate the free energy Eq.~\ref{Eq:Free energy_defects_simple} and compare it to the one with an additional new defect (local condition). Once the latter one is favorable, the new defect is added.

\begin{figure}
	\centering
	\includegraphics[width=\linewidth]{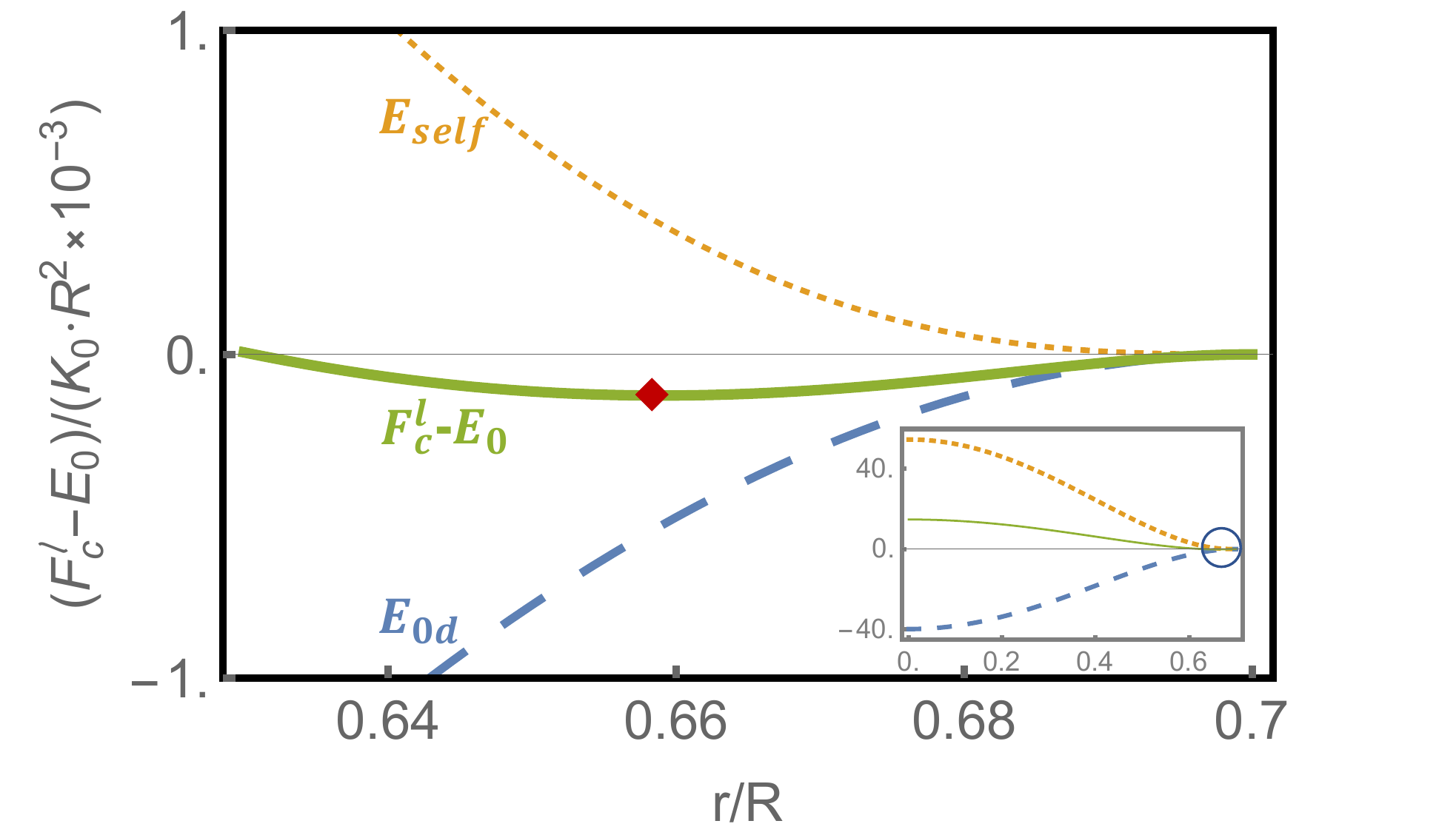}
	\caption{\footnotesize 
		The 1D energy plot for the first disclination: The dotted line corresponds to disclination self-energy $E_{self}$  (Eq.~\ref{Eq:dd_term}), the dashed line to the Gaussian curvature-disclination interactions $E_{0d}$ and the solid line is the result of the addition of both energies $F_c^l-E_0$ (Eq.~\ref{Eq:Free energy_defects_simple}) as a function of the location of disclination in the shell for $\theta_m=0.7$. The energy goes through a minimum for $r=0.66$. The inset graph shows the zoom-out energy plot where the circle region corresponds to the main graph.
		}\label{firstdisc}
\end{figure}

For small values of $\theta_m$, the cap grows defect free. In Fig.~\ref{firstdisc} we plot the energy of
a spherical cap for $\theta_m=0.7$. The dotted line in the figure shows the disclination self-energy $E_{self}$, the dashed line the Gaussian curvature-disclination interactions $E_{0d}$ and the solid line is the sum of both energies as a function of the location of disclination in the cap. The diamond in the figure corresponds to the minimum of energy and indicates the location of the first (and only) disclination appearing in the cap, around $r\sim0.66$. This value is very close to the geodesic distance following from the local ``screening'' of the Gaussian curvature 
%\begin{equation}\label{Eq:screen}
$\int d^2{\bf x}~s({\bf x}) = \int d^2{\bf x}~K({\bf x}) \rightarrow \frac{\pi}{3}=2 \pi (1-\cos(\theta_m))$
%\end{equation}
such that $r = \arccos(5/6) = 0.59$. Somewhat counter intuitively, the first disclination does not appear at the center of the cap, which is the result of the competition between the disclination self-energy whose minimum is at the boundary and the Gaussian curvature-disclination interaction $E_{0d}$ with its minimum energy occurring at the cap center, see Fig.~S2 in SI Appendix where the contourplots of the different elastic free energies as a function of the location of the first disclination, $r$ are shown. As the shell grows, the appearance of a new disclination becomes energetically favorable, $i.e.$ a new energy valley for the formation of a new disclination emerges, as illustrated in Fig.~\ref{largeT}b, where we show the contour plots of total elastic energies for spherical caps with $\theta_m=0.8$ through $\theta_m=\pi$. The bigger ball in each plot indicates the position of the latest energy well, which is where the addition of the next disclination takes place. Remarkably, both in the continuum model and simulations, during the growth process, the disclinations always appear in the positions that eventually become the vertices of an icosahedron. 

Results with the discrete model Eq.~\ref{Eq:discrete} are shown in Fig.~\ref{largeT}a. Here again, the disclinations universally appear at the vertices of an icosahedron, in complete agreement with the analytical calculation. The simulations were performed for all values between $T=7$ and $T=21$ and in all cases the IO was achieved without a single error. The size of the core in Fig.~\ref{largeT}a is commensurate with $T=13$ structures. We note that for these simulations the proteins spontaneous radius $1/H_0$ is much smaller than the core radius, $R_c$ ($R_c H_0 \gg 1$), a point that is discussed in more detail further below. In SI Appendix we provide a movie illustrating the growth of a $T=21$ structure, which includes 420 triangles.

\subsection*{The role of stretching and bending rigidity}\label{Sect:res_stretch}

\begin{figure}
	{\includegraphics[width=\linewidth]{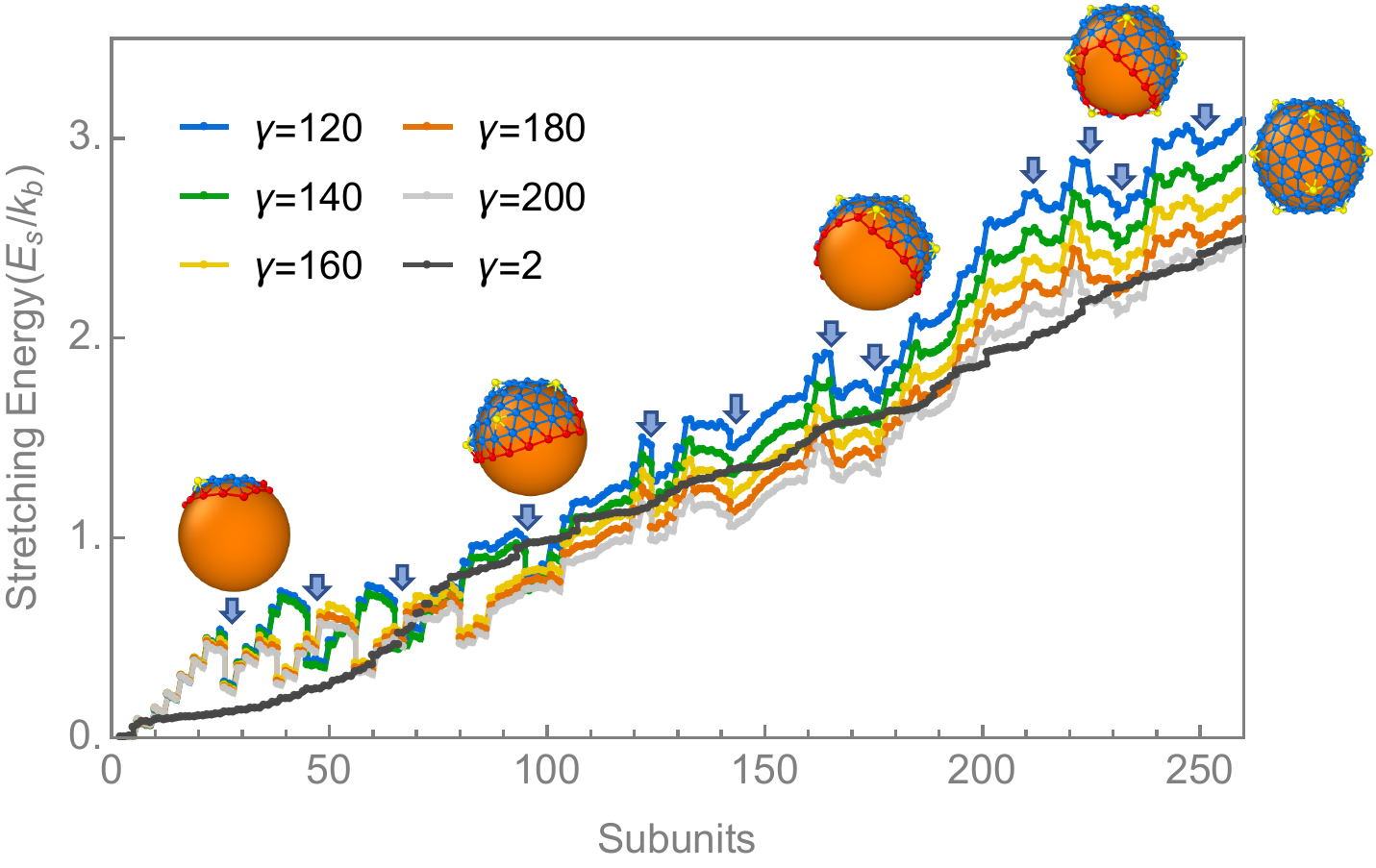}} \\
	\caption{\footnotesize The stretching energy of a T=13 shell as a function of number of trimeric subunits. For small FvK numbers ($\gamma =2$, black line), there is no significant drop in energy as a pentamer forms.  However, for large FvK numbers ($\gamma \gg 1$), the formation of pentamers drastically lowers the energy of the elastic shell. }
	\label{shellenergy}
\end{figure}

Figure~\ref{shellenergy} shows the stretching energy vs $N$ (number of subunits assembled) as a $T=13$ shell grows for six different values of FvK $\gamma > 1$. We note that for large spontaneous radius of curvature and small $\gamma$ when bending rigidity is dominant, no large icosahedral shell assembles successfully. Rather interestingly, there are conspicuous differences in the dynamics as a function of the FvK parameter $\gamma$.

For small values of $\gamma = 2$ (thick black line in Fig.~\ref{shellenergy}) the shell elastic energy grows almost linearly as a function of $N$ but does not show IO. This takes place for higher $\gamma$-values. The arrows in Fig.~\ref{shellenergy} indicate a drop in the elastic energy associated with the appearance of pentamers, see SI Appendix for more details.  At the beginning of the growth, the shells with different values of $\gamma$ might follow different pathways and thus, the number of hexamers vary before the first few pentamers form. However, as the shell grows, the pentamers appear precisely at the same place, independently of $\gamma$. Note that the bending energy of the shells always grows linearly as a function of number of subunits for any $\gamma$ (see SI Appendix). 

\section*{Discussion}\label{Sect:Conclusions}

Our results show that for large shells ($T>4$) successful assembly into IO requires a non-specific attractive interaction between protein subunits and a template. This interaction is implicit in the continuum model and is included as a generic attractive Lennard-Jones potential in the simulations. Furthermore, we find that the location of pentamers are completely controlled by the stretching energy as it is the case in the continuum elasticity theory.  

In the absence of the template, small spherical crystals ($T=1$ and $T=3$) assemble spontaneously, for almost any FvK parameter $\gamma$. However, as we increase the spontaneous radius of curvature, the final structure depends on the value of $\gamma$. For small $\gamma$, large spherical shells without any specific symmetry form, and at large $\gamma>5$ curved hexagonal sheets, which eventually assemble into tubular or conical structures are obtained. Thus our results predict that large shells with IO cannot grow without template.

A template can have a significant impact on the structure and symmetry of the shell. While a weak subunit-core attractive interaction has a minimal role in the shell shape, a very strong subunit-core interaction will override the mechanical properties of proteins. The subunits sit tightly on the template to form a sphere with no specific symmetry. We were able to observe large shells with IO only for $\eta\sim1$ but at high $\gamma$. In this regime, in order for pentamers to overcome the core attraction and form in the ``correct'' position, they must assume a symmetric shape and buckle up (see Fig.~\ref{gallery}b). Indeed a strong bending energy is needed to overcome the shell adsorption. We find that without decreasing $\gamma$ (increasing $k_b$) but with increasing spontaneous curvature, the bending energy associated with the deviation from the preferred curvature of subunits adsorbed to the core becomes strong enough to make the pentamers buckle and assume a smooth shape. Quite interestingly, we find that this is the strategy that the nature has taken to form large shells with IO.

The role of the inner core or the preformed scaffold layer presented above is very similar to the role of SPs, which assemble at the same time as the capsid proteins (CPs), {\it i.e.}, the template grows simultaneously with the capsid (see Fig.~\ref{cartoon}). In fact one can think of the inner core as a permanent ``inner scaffold''~\cite{Coulibaly2005}. For example, Bacteriophage P22 has a T=7 structure, but in the absence of scaffolding (Fig.~\ref{gallery}, P22) often a smaller $T=4$ forms. Similarly, Herpesvirus makes a $T=16$ structure but without the SPs, a $T=7$ assembles. More relevant to the present study is the case of Infectious Bursal Disease Virus (IBDV) a dsRNA virus that in the presence of SPs forms a $T=13$ capsid but in the absence the subunits assemble to form a $T=1$ capsid. This is exactly the condition for formation of the $T=13$ structure in Fig.~\ref{largeT} where the preferred curvature between subunits is such that in the absence of scaffold they form a $T=1$ structure. Reoviridae virus family also form $T=13$ but they have multi-shell structures, which act as inner cores. For instance, in this family Bluetongue virus is a double capsid particle, outer (necessary for infection) and inner capsid (encloses RNA genome). The inner capsid, termed as ``core'' has two protein layers. The surface layer (or shell) is a $T = 13$ capsid that assembles around the inner shell, a $T = 2$ structure (an inner core). Interestingly, it has been suggested that there is an evolutionary connection between SPs of IBDV and inner capsid of Bluetongue virus~\cite{Coulibaly2005}.

\section*{Conclusions}

Our model establishes that successful self-assembly of components into a spherical capsid with IO requires a template that determines the radius of the final structure. This template is very non-specific, and in its absence, protein subunits assemble into either smaller capsids or structures without IO. 
\begin{figure}
	{\includegraphics[width=\linewidth]{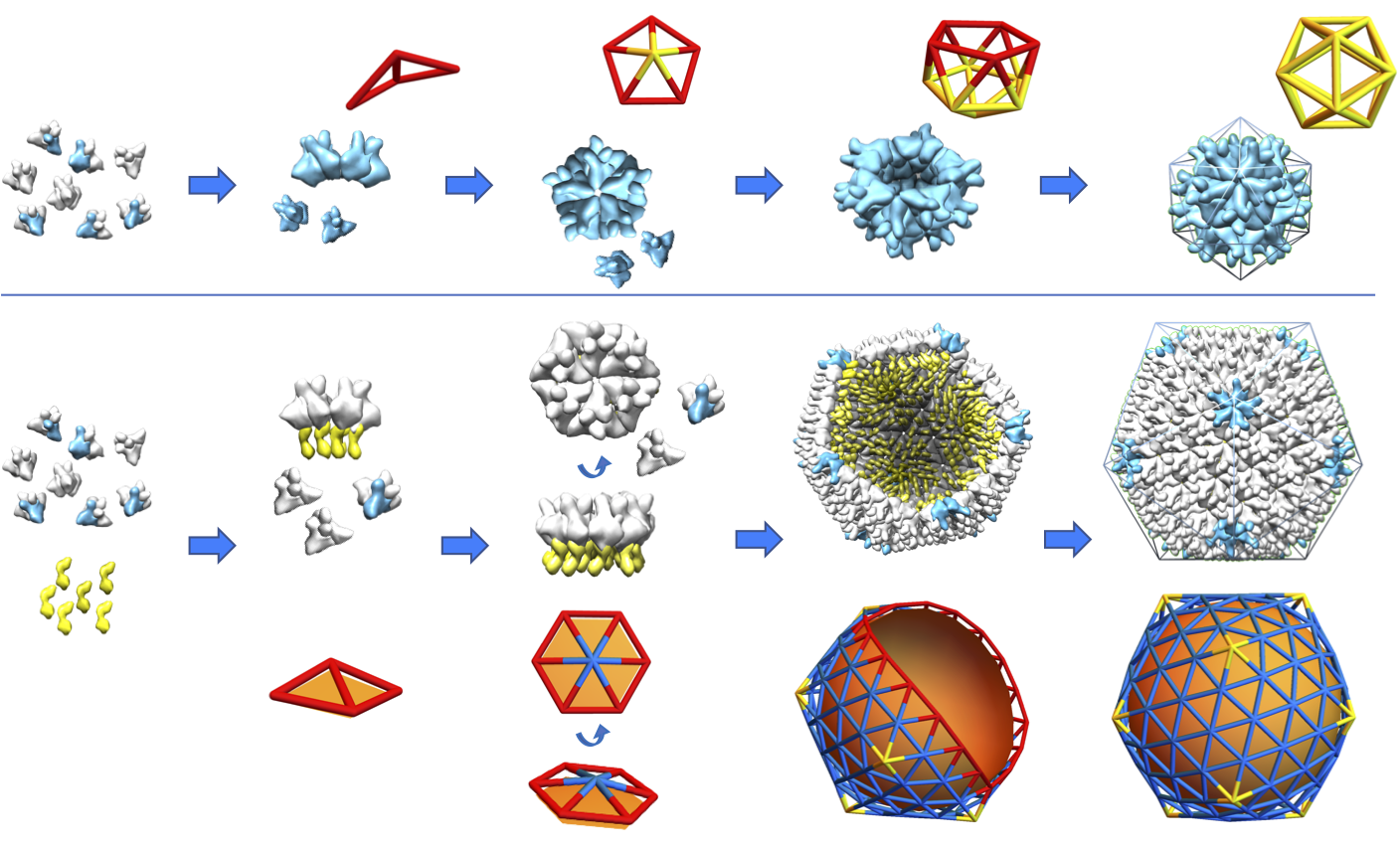}} \\
	\caption{\footnotesize The role of scaffolding proteins (SPs) in the formation of $T=13$ capsid of IBDV. Without SPs, the CPs (blue and white subunits) of IBDV form $T=1$ structure (upper figure). In the presence of SPs (yellow subunits), they form $T=13$ structure (lower figure).  The results of our simulations are also illustrated next to each intermediate step. Note that SPs (yellow subunits) do not assemble without the CPs but probably experience some conformational changes during the assembly.   However, our focus here is solely on the impact of scaffolding on the CPs resulting in a change in the capsid T number. For preformed SPs, like in the case of bluetongue virus, the core is spherical and there is no indication of any changes on the size of spherical template during the assembly.}
	\label{cartoon}
\end{figure}

Even though the focus of the above study was on the impact of the preformed scaffolding layer, based on the experimental observations we conclude that the SPs, which assemble simultaneously with CPs (Fig.~\ref{cartoon}), play basically the same role as the inner core in the assembly of large icosahedral shells. Figure \ref{cartoon} shows that in the absence of SPs, CPs of IBDV form a $T=1$ structure but when the same IBDV proteins co-assemble with SPs (yellow units) a $T=13$ forms. The figure also shows the pathway of formation of $T=1$ and $T=13$ structures obtained in our simulations. We emphasize that the mechanical properties of subunits are the same for both shells, the difference in structures arises from the substrate or SPs.

The contribution of the SPs is twofold. The CPs of many viruses including bluetongue virus noted above do not assemble in the absence of SPs. On the one hand, it appears that SPs lower the energy barrier and help capsid subunits to aggregate. On the other hand, by forcing the CPs to assemble into a structure larger than their spontaneous radius of curvature, they contribute to preserving IO.

Examples of the role of templates on the formation of spherical crystals are not limited to viruses, but include crystallization of metals on nanoparticles~\cite{Huixin2011}, solid domains on vesicles~\cite{Korlach1999,GuttmanSapirSchultzButenkoOckoDeutschSloutskin2016}, filament bundles~\cite{Grason2012} and colloidal assemblies at water-oil interfaces~\cite{BauschMe2003}.
Nevertheless, it has been shown~\cite{Meng2014} that sufficiently rigid crystals grow as almost flat sheets free of defects, unable to assemble with IO. This regime, however, seems not to be accessible to viral capsids, as the hydrophobic interaction between monomers force close-packing structures that are incompatible with grain boundaries.

This study shed light at fundamental scale on the role of mechanical properties of building blocks and scaffolding proteins. The proposed mechanism is consistent with available experiments on viruses involving either scaffolding proteins or inner capsids. Further experiments will be necessary to validate many predictions of our described mechanism.

\section*{acknowledgements}
The authors would like to thank Greg Grason for many helpful discussions. The S.L. and R.Z. work were supported by NSF Grant No. DMR-1719550 and A.T. by NSF Grant No. DMR-1606336. PR is funded through Senior Investigator Award, Welcome Trust under Grant No: 100218. A.T. and R.Z. 
thank the Aspen Center for Physics where part of this work was done with the support of the NSF Grant No.PHY-1607611.

% Bibliography
\bibliography{bibfile}
\end{document}

\title{
Why large icosahedral viruses need scaffolding proteins: The interplay of Gaussian curvature and disclination interactions.
Supplementary Information}
\author{Siyu Li}
\affiliation{Department of Physics and Astronomy,
   University of California, Riverside, California 92521, USA}
\author{Polly Roy}   
\affiliation{Department of Pathogen Molecular Biology, Faculty of Infectious and Tropical Diseases, London School of Hygiene and Tropical Medicine, London WC1E 7HT, UK}  
\author{Alex Travesset}
\affiliation{Department of Physics and Astronomy, Iowa State and Ames Lab, Ames, Iowa 50011-3160, USA}
\author{Roya Zandi}
\affiliation{Department of Physics and Astronomy,
   University of California, Riverside, California 92521, USA}

\maketitle
\section*{Elastic Energy in Continuum Model}
\begin{figure}[H]
\centering
\includegraphics[width=0.5\linewidth]{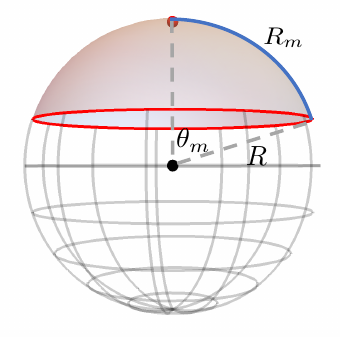}
\caption{a spherical cap with geodesic size $R_m$ and radius $R$, the cap center is marked as a red dot on the top.}
\label{cap}
\end{figure}

In this section we provide the detailed calculations for the elastic energy term($F_c^l$) in continuum model, Eqs.~10 and 11 given in the text. We define the elastic free energy as
\begin{equation}\label{F1}
F_c^l=\frac{1}{2 K_0}\int d^2{\bf x} \sqrt{g}\left(\Delta\chi\right)^2 \ ,
\end{equation}
where $g_{\mu \nu}$ is the metric of the surface. The Laplacian($\Delta$) is then defined as $-\frac{1}{\sqrt{g}}\partial_{\mu} g^{\mu \nu}\partial_{\nu}$. For a spherical surface (Fig.~\ref{cap}), 
% where $R_m=\theta_m \cdot R$ as the cap geodesic size , $\theta_m$ as the edge polar angle and $R$ correspond to the sphere radius, 
the metric tensor and the laplacian operator are $ds^2=dr^2+R^2\sin^2(\frac{r}{R})d\phi^2$ and $\Delta=\frac{1}{\sin(\frac{r}{R})}\partial _r\left(\sin(\frac{r}{R})\partial _r\right)+\frac{1}{R^2\sin^2(\frac{r}{R})}\partial _{\phi \phi }$, respectively. The Airy Stress function $\chi$ satisfies 
\begin{equation}\label{Eq:bieq}
\frac{1}{K_0}\Delta ^2\chi({\bf x}) =  s({\bf x})- K({\bf x}) .
\end{equation}
with $s({\bf x})=\sum _{i=1}^N\frac{q_i}{\sqrt{g}}\delta ({\bf x}-{\bf x}_i)$ the disclination density and $K({\bf x})=\frac{1}{R^2}$ gaussian curvature.  We seek the solutions of $\chi$ subject to the stress free boundary conditions, $\sigma ^{rr}=\sigma ^{\phi r}=0$. Following the defination of the stress tensor, the boundary conditions are then
\begin{align}
\sigma ^{rr}&=\frac{1}{R}\cot(\frac{r}{R}) \partial_r\chi+\frac{1}{R^2\sin^2(\frac{r}{R})}\partial_{\phi\phi}\chi  =0 \label{bcrr}\\
\sigma ^{\phi r}&=-\frac{1}{R\sin(\frac{r}{R})}\partial _r\left(\frac{1}{R\sin(\frac{r}{R})} \partial_{\phi}\chi \right) =0 \label{bcrph}
\end{align} 
at $r=R\theta_m=R_m$. \\
% the green function $G({\bf x},{\bf y})$
% then integrate the free energy $F_c^l$ by parts twice with subtitution of green function to get
Constructing the green function 
\begin{equation}\label{eqG2}
\Delta^2 G({\bf x},{\bf x}_i) = \frac{1}{\sqrt{g}}\delta({\bf x}-{\bf x}_i),
\end{equation}
subject to the dirichlet  ($G(R_m,r_i)=0$) and stress free boundary conditions given in Eqs.~(\ref{bcrr}) and (\ref{bcrph}), we find the airy stress function can be written as the integral of the green function with the source field
\begin{align}\label{chi0}
% \chi(r) &= K_0 \int d^2r' ~G(r,r') \left( \sum _{i=1}^N\frac{q_i}{2\pi }\frac{1}{\sin[r']}\delta (r'-r_i)-\frac{1}{R^2} \right) \nonumber\\
\chi({\bf x}) &= K_0 \int d^2 {\bf x'} ~G({\bf x},{\bf x'}) \left( s({\bf x'})-K({\bf x'}) \right) \nonumber\\
&=K_0 R^2 \left(\sum_{i=1}^N q_i G({\bf x},{\bf x}_i) -  \int d^2 {\bf x'} ~G({\bf x},{\bf x'}) \right)  \nonumber\\
&=K_0 R^2 \left(\sum_{i=1}^N q_i G({\bf x},{\bf x}_i) -  \Omega({\bf x}) \right), 
\end{align}
where {\bf x} has been rescaled in the unit of R and $\Omega({\bf x}) \equiv \int d^2 {\bf x'}~G({\bf x},{\bf x'})$. Thus we have
\begin{equation}\label{eqomega}
\Delta ^2\Omega({\bf x})=1. 
\end{equation}
Integrating the free energy in Eq.~(\ref{F1}) by parts and using  the green function $G$, we find
\begin{align}\label{F2}
F_c^l &= \frac{K_0 R^2}{2}\int d^2{\bf x} d^2{\bf y} \left[(K({\bf x})-s({\bf x}) )G({\bf x},{\bf y}) (K({\bf y})-s({\bf y})) \right] \nonumber\\
% \right. \times \nonumber\\
% & \hspace{4cm}\left. 
% &=K_0 R^2\left(\sum_{i,j}^N s^2 G(r_i,r_j)-\sum_i^N 2 s\int d^2r G(r,r_i) \right. \nonumber\\
% &\hspace{4cm}\left.+ \int d^2r d^2 r'G(r,r')\right)
&=\frac{K_0 R^2}{2}\left(\int d^2 {\bf x} ~\Omega({\bf x})-\sum_{i=1}^N 2 q_i ~\Omega({\bf x}_i)+\sum_{i=1}^N \sum_{j=1}^N q_i q_j G({\bf x}_i,{\bf x}_j)\right)  \nonumber \\
&=E_0(\theta_m) + \sum_{i=1}^N q_i \hat{E}_{0d}({\bf x}_i, \theta_m) + \sum_{i=1}^N \sum_{j=1}^N q_i q_j \hat{E}_{dd}({\bf x}_i,{\bf x}_j, \theta_m)  \nonumber\\
% &=E_0(\theta_m) + \sum_{i=1}^N q_i E_{0d}({\bf x}_i, \theta_m) + \sum_{i=1}^N E_{self}({\bf x}_i) \nonumber\\
% &\hspace{3cm}+ \sum_{i=1}^N \sum_{j>i}^N q_i q_j \hat{E}_{dd}({\bf x}_i, {\bf x}_j) 
&=E_0(\theta_m) + \sum_{i=1}^N E_{0d}({\bf x}_i, \theta_m) + \sum_{i=1}^N E_{self}({\bf x}_i) + \sum_{i=1}^N \sum_{j>i}^N E_{dd}({\bf x}_i, {\bf x}_j) 
\end{align}\\
% \begin{align}\label{Eq:dd_term}
% F_c^{dd} &= \sum_{i=1}^N \sum_{j=1}^N q_i q_j E_{dd}({\bf x}_i,{\bf x}_j, \theta_m)
% \nonumber\\
% &=\sum_{i=1}^{i=N} q_i^2 E_{dd}({\bf x}_i,{\bf x}_i) + \sum_{i=1}^N \sum_{j>i}^N q_i q_j \hat{E}_{dd}({\bf x}_1, {\bf x}_2) \nonumber \\
% &\equiv \sum_{i=1}^N E_{self}({\bf x}_i) + \sum_{i=1}^N \sum_{j>i}^N q_i q_j \hat{E}_{dd}({\bf x}_1, {\bf x}_2),
% \end{align}
where the first term in Eq.~(\ref{F2}) describes the gaussian curvature self energy, the second term corresponds to disclination-gaussian curvature interaction, the third term denotes disclination self-energy and the last term is the disclination-disclination interaction energy.  In the next section we solve the green function $G({\bf x}_i,{\bf x}_j)$ and $\Omega({\bf x})$ for when there exists only one disclination at the center of a spherical cap and then solve it for when it is placed off center.

\subsubsection*{Center Disclination}\label{center}
We solve Eqs.~(\ref{eqG2})-(\ref{eqomega}) for a radially symmetric system when we only have one disclination at the center of a spherical cap($r_i=0$). We find
\begin{eqnarray}
2\pi G(r,0)&=&\text{Li}_2(-\tan^2(\frac{r}{2}))+\log(1+\tan^2(\frac{r}{2}))\log(\tan(\frac{r}{2}))+u  \label{G2c}\\
\Omega(r)&=&\text{Li}_2(-\tan^2(\frac{r}{2}))+\frac{1}{2}\log(1+\tan^2(\frac{r}{2}))^2+u  \label{Omegac}
\end{eqnarray}
where $\text{Li}_2$ is a polylogarithm $\text{Li}_2 (x)=\sum_{n=1}^{\infty}\frac{x^n}{n^2}$ and $u$, the homogeneous solution of biharmonic equation $\Delta^2 u=0$, is given by
\begin{equation}
u=C+D\log(1+\tan^2(\frac{r}{2}))
\end{equation}
% and assume $\Delta \chi _{ds}= \Gamma_{ds}$, $\Delta \chi _{g}= \Gamma_{g}$ the equations become:
% \begin{align}
% \Delta \Gamma_{ds}&=\frac{1}{\sin[r]}\partial _r\left(\sin[r]\partial _r\Gamma_{ds}\right)
% =\frac{1}{2\pi}\frac{1}{\sin[r]}\delta(r) \\
% \Delta\Gamma_g&=\frac{1}{\sin[r]}\partial _r\left(\sin[r]\partial _r\Gamma_g\right)=1
% \end{align}
% \begin{align}
% &\sigma_{ds}=\frac{1}{2\pi}\log(\tan(\frac{r}{2}))+u \\
% &\sigma_g=\log(1+\tan^2(\frac{r}{2}))+u 
% \end{align}
Applying the boundary conditions Eqs.~(\ref{bcrr})-(\ref{bcrph}), $G(\theta_m,0)=0$ and $\Omega(\theta_m)=0$, we find 
% \begin{widetext}
\begin{align}
2\pi G(r,0)=&\text{Li}_2(-\tan^2(\frac{r}{2}))+\log(1+\tan^2(\frac{r}{2}))\log(\tan(\frac{r}{2}))\nonumber\\
&\hspace{2cm}+2\log(\cos(\frac{r}{2}))\left(\cot^2(\frac{\theta_m}{2}) \log(\cos(\frac{\theta_m}{2}))+\log(\sin(\frac{\theta_m}{2}))) \right)\nonumber\\
&-\text{Li}_2(-\tan^2(\frac{\theta_m}{2}))-\log(1+\tan^2(\frac{\theta_m}{2}))\log(\tan(\frac{\theta_m}{2}))\nonumber\\
&\hspace{2cm}-2 \log(\cos(\frac{\theta_m}{2})\left(\cot^2(\frac{\theta_m}{2}) \log(\cos(\frac{\theta_m}{2}))+ \log(\sin(\frac{\theta_m}{2}))\right)
\end{align}
and 
\begin{align} \label{eq:omega}
\Omega(r)=&\text{Li}_2(-\tan(\frac{r}{2})^2)+\frac{1}{2} \log(1+\tan^2(\frac{r}{2}))^2 \nonumber\\
&\hspace{2cm}+\log(1+\tan^2(\frac{r}{2}))
\cot^2(\frac{\theta_m}{2}) \log(1+\tan^2(\frac{\theta_m}{2})) \nonumber\\
&-\text{Li}_2(-\tan(\frac{\theta_m}{2})^2)-\frac{1}{2} \log(1+\tan^2(\frac{\theta_m}{2}))^2 \nonumber\\
&\hspace{2cm}-\log(1+\tan^2(\frac{\theta_m}{2}))
\cot^2(\frac{\theta_m}{2}) \log(1+\tan^2(\frac{\theta_m}{2}))
\end{align}
We will employ the expressions for $G(r,0)$ and $\Omega(r)$ to calculate the free energy of the system given in Methods section.
% \end{widetext}
% the free energy from equation~(\ref{F2}) is therefore
% \begin{align}
% &F=Y\left(\int d^2r ~\Omega(r)-\Omega(0)+s^2G(0,0)\right) \label{Fc}
% \end{align}
% and the whole stress we get from $\Delta \chi$ is thus
% \begin{align}
% &\frac{2\pi}{Y s}\sigma_{ds}= \Delta G(r,0) \nonumber\\
% &=\log(\tan(\frac{r}{2}))-\left(\cot^2(\frac{\theta_m}{2}) \log(\cos(\frac{\theta_m}{2}))+\log(\sin(\frac{\theta_m}{2}))\right) \label{stressdsc}\\
% &\frac{1}{Y}\sigma _g=\Delta \Omega(r) \nonumber\\
% &=-1-\log(\cos^2(\frac{r}{2}))+\cot(\frac{\theta
% c}{2})^2 \log(\sec^2(\frac{\theta c}{2})) \label{stressg}
% \end{align}
\subsubsection*{Off-center disclination}\label{offcenter}
We now consider an off-center disclination. In this case, the gaussian curvature contribution $\Omega(r)$ remains as in the previous section but the green function $G({\bf x},{\bf x}_i)$ can be obtained using the function $\Gamma$ as follows,
\begin{eqnarray}
\Delta \Gamma({\bf x},{\bf x}_i) &=& \frac{1}{\sqrt{g}}\delta({\bf x}-{\bf x}_i) \label{eqG}\\
\Delta G({\bf x},{\bf x}_i)&=&\Gamma({\bf x},{\bf x}_i). \label{eqG22}
\end{eqnarray}
Expressing $\Gamma({\bf x},{\bf x}_i)$ as the sum of eigenstates, we have
\begin{equation} \label{G}
2\pi \Gamma({\bf x},{\bf x}_i)=\log(\tan(\frac{r_>}{2}))-\sum_{n=1}^{\infty} \frac{1}{n}\frac{\tan^n(\frac{r_<}{2})}{\tan^n(\frac{r_>}{2})}\cos(n\phi),
\end{equation}
which we insert in Eq.~(\ref{eqG22}). Then we find

\begin{align}\label{G2}
% 2\pi G(r,r_i)
% &=\text{Li}_2 (-\tan^2(\frac{r_>}{2}))+\log(\tan(\frac{r_>}{2}))\left(\log(1+\tan^2(\frac{r_>}{2}))+\log(1+\tan^2(\frac{r_<}{2})\right) \nonumber\\
% &~~~+\sum _{n=1}^{\infty}\frac{1}{n}\left(\frac{1}{n}\frac{\tan^n(\frac{r_<}{2})}{\tan^n(\frac{r_>}{2})} -H_n(\cot(\frac{r_>}{2}))\tan(\frac{r_<}{2}){}^n -\cot^n(\frac{r_>}{2}) H_n(\tan(\frac{r_<}{2}))\right)
% \cos[n\phi] +U(r) \label{G2}
2\pi G({\bf x},{\bf x}_i)=&\text{Li}_2 (-\tan^2(\frac{r_>}{2})) +\log(\tan(\frac{r_>}{2}))\left(\log(1+\tan^2(\frac{r_>}{2}))+\log(1+\tan^2(\frac{r_<}{2})\right) \nonumber\\
&+\sum _{n=1}^{\infty}\frac{1}{n}\left(\frac{1}{n}\frac{\tan^n(\frac{r_<}{2})}{\tan^n(\frac{r_>}{2})} -H_n(\cot(\frac{r_>}{2}))\tan^n(\frac{r_<}{2}) -\cot^n(\frac{r_>}{2}) H_n(\tan(\frac{r_<}{2}))\right)
\cos(n\phi) \nonumber\\
&+U({\bf x}) 
\end{align}
where $r_<$ and $r_>$ correspond to min(r,$r_i$) and max(r,$r_i$) respectively. The function $U({\bf x})$ denotes the homogenous solution of biharmonic($\Delta^2$) operator,
\begin{equation}\label{U}
U({\bf x})=c_0+d_0\log(1+\tan(\frac{r}{2})^2) + \sum_{n=1}^{\infty}\frac{1}{n}\left(c_n \tan(\frac{r}{2})^n+ d_n H_n(\tan(\frac{r}{2}))\right)\cos(n\phi)
\end{equation}
with $H_n$ the basis function of the biharmonic operator that is given by
\begin{equation} 
H_n(x)=\frac{x^{2+n}}{n} \left(\frac{1}{1+x^2}-\frac{1}{1+n } {}_2F_1 \left(2,1+n,2+n,-x^2\right)\right) \label{H},
\end{equation}
where ${}_2F_1$ is the hypergeometric function with $\frac{1}{1+n} {}_2F_1(2,1+n,2+n,x)=\sum_{i=0}^{\infty}\frac{i+1}{n+i+1} x^i$. Applying the boundary conditions Eqs.~(\ref{bcrr})-(\ref{bcrph}) and $G(\theta_m,r_i)=0$, we obtain the coefficients in Eq.~(\ref{U}) as
\begin{eqnarray}\label{coefs}
d_0&=&-\log(b)+\frac{1}{2} \left(1+b^{-2}\right) \log\left(\frac{1+b^2}{1+x \rho^2}\right) \nonumber\\
c_0&=&-\text{Li}_2(-b^2)-\log(b)\left(\log(1+b^2)+\log(1+x\rho^2)\right) -d_0 \log(1+b^2) \nonumber\\
d_n&=&\frac{1- \left(1+b^{-2}\right)x \rho^{-n}n H_n(x \rho)}{1-\left(1+b^{-2}\right) b^{-n}nH_n(b)} \left(\frac{x \rho}{b}\right)^nb^{-n} \nonumber\\
c_n&=&\frac{b^{-2n}}{n \big(b^{2+n}-(1+b^2) n H_n(b)\big)} \bigg({x\rho}^n \Big(-b^{2+n}+n H_n(b)
+ b^n n H_n(b^{-1})\left(b^{2+n}-(1+b^2) n H_n(b)\right)\Big)\bigg.\nonumber\\
&&\hspace{5cm}\bigg.+b^{2+n} n H_n(x\rho)\bigg) \nonumber\\
% c_n&=\frac{b^{-2n}}{n^2 (b^{2+n}-(1+b^2) n H_n(b)} \Big( -{x\rho}^n b^{2+n}+{x\rho}^n n H_n(b)  \nonumber\\
% &+{x\rho}^n b^n n H_n(b^{-1})\left(b^{2+n}-(1+b^2) n H_n(b)\right)+b^{2+n} n H_n(x\rho)\Big)
\end{eqnarray}
where $b=\tan(\frac{\theta_m}{2})$ and $x \rho=\tan(\frac{r_i}{2})$.

% Following equation~(\ref{F2}),~(\ref{Omegac}),~(\ref{G2}) the elastic energy of the shell as well as the stress caused by one disclination located at $r_i$ are
% \begin{align}
% \frac{2\pi}{Ys}\sigma_{ds}(r,r_i)&=\Delta G(r,r_i) \nonumber\\
% &=d_0+\log(\tan(\frac{r_>}{2})) \nonumber\\
% &-\sum _n^{\infty}\frac{1}{n}\left(
% \frac{\tan^n(\frac{r_<}{2})}{\tan^n(\frac{r_>}{2})} - d_n\tan(\frac{r}{2})^n\right)\cos(n\phi) \label{stressds}
% \end{align}

% When $\theta_m$ and $\rho $ is small, this stress retrieve the same form as in the planar coordinates.

% $\sigma_{\text{grason}}=$ \\

% \(
% \hspace{-0.8cm}
% \left\{ 
% \begin{array}{ll}
% \frac{1}{2}-\frac{ \rho ^2}{2R^2}+ \log\left[\frac{\rho }{R}\right]\\
% \;+\sum _{n=1}^{\infty }\frac{1}{n}\left(-\left(\frac{r}{\rho }\right)^n+\frac{n(1+n)}{R^{n+2}}\left(\frac{R^2}{n}-\frac{\rho^2}{1+n}\right)\left(\frac{\rho }{R}\right)^n r^n\right)\cos[n \phi ]\\
% \hspace{7cm}r<\rho  \\
% \frac{1}{2}-\frac{ \rho ^2}{2R^2}+\log\left[\frac{r}{R}\right]\\
% \;+\sum _{n=1}^{\infty }\frac{1}{n}\left(-\left(\frac{\rho }{r}\right)^n+ \frac{n(1+n)}{R^{n+2}}\left(\frac{R^2}{n}-\frac{\rho ^2}{1+n}\right)\left(\frac{\rho }{R}\right)^nr^n\right)\cos[n \phi ] \\
% \hspace{7cm} r>\rho  \\
% \end{array}
% \right.
% \)
% \\
% \\

% The figure shows the deviation of our stress and grason{'}s stress when $\theta c=\frac{\pi}{3},~\rho=0.2\pi$ \\

% \begin{figure}[H]
% \includegraphics[width=0.8\linewidth]{comparestress.pdf}
% \caption{\footnotesize Comparison between $\sigma_{geodesic}$ and $\sigma_{grason}$}
% \label{compare}
% \end{figure}

\section*{Energy ContourPlots}

\begin{figure}
\centering
\subfloat[$\text{E}_{0d}$, equation~(\ref{F2})]{\includegraphics[width=0.5\linewidth]{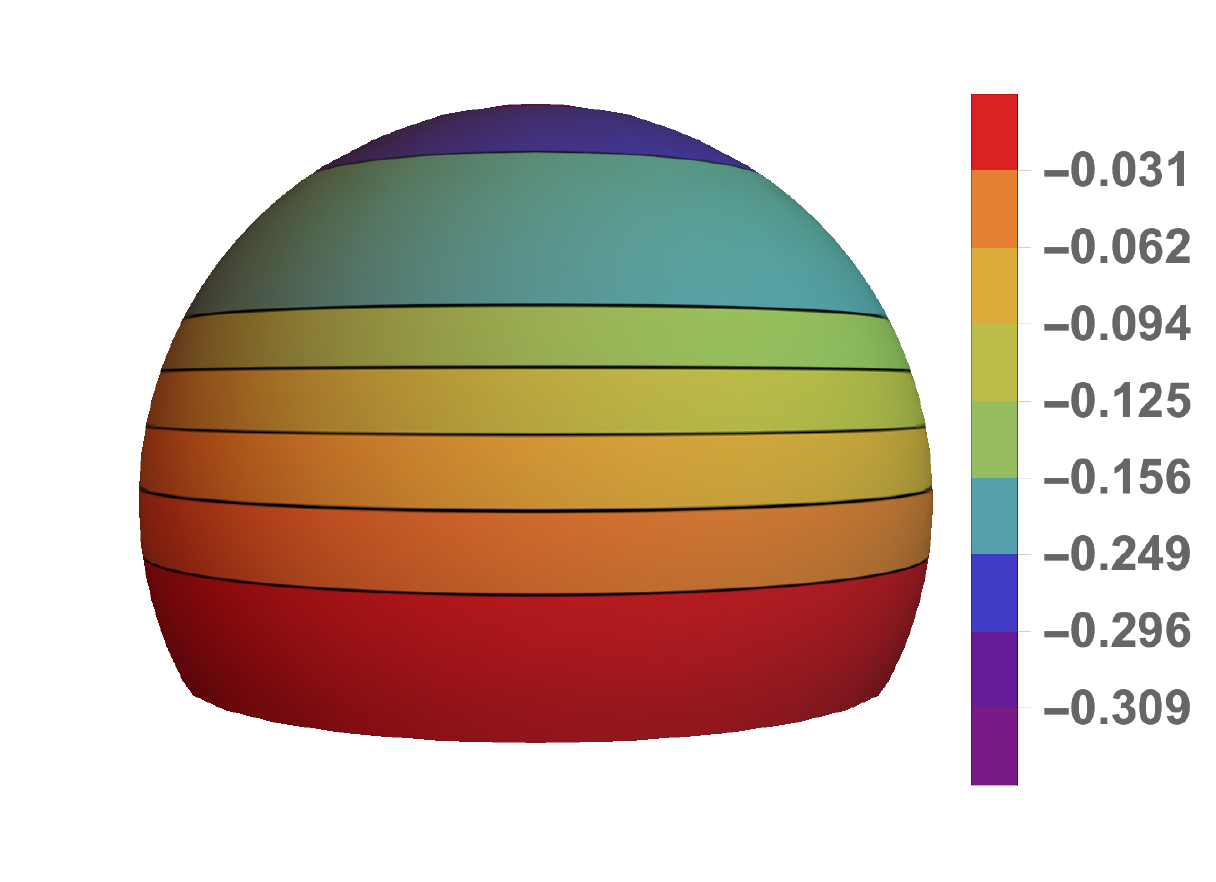}}\\
\subfloat[$\text{E}_{self}$, equation~(\ref{F2})]{\includegraphics[width=0.5\linewidth]{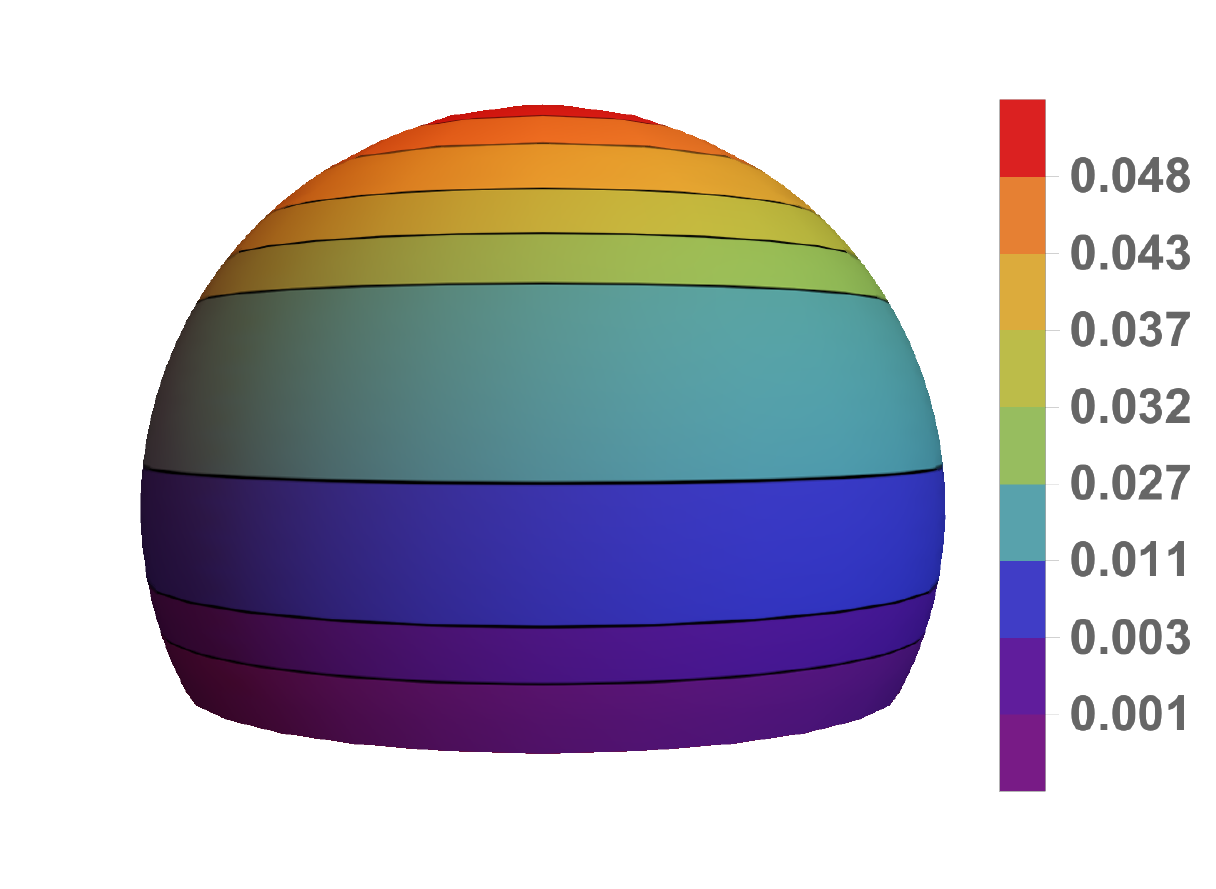}}\\
\subfloat[$\text{E}_{dd}$, equation~(\ref{F2})]{\includegraphics[width=0.5\linewidth]{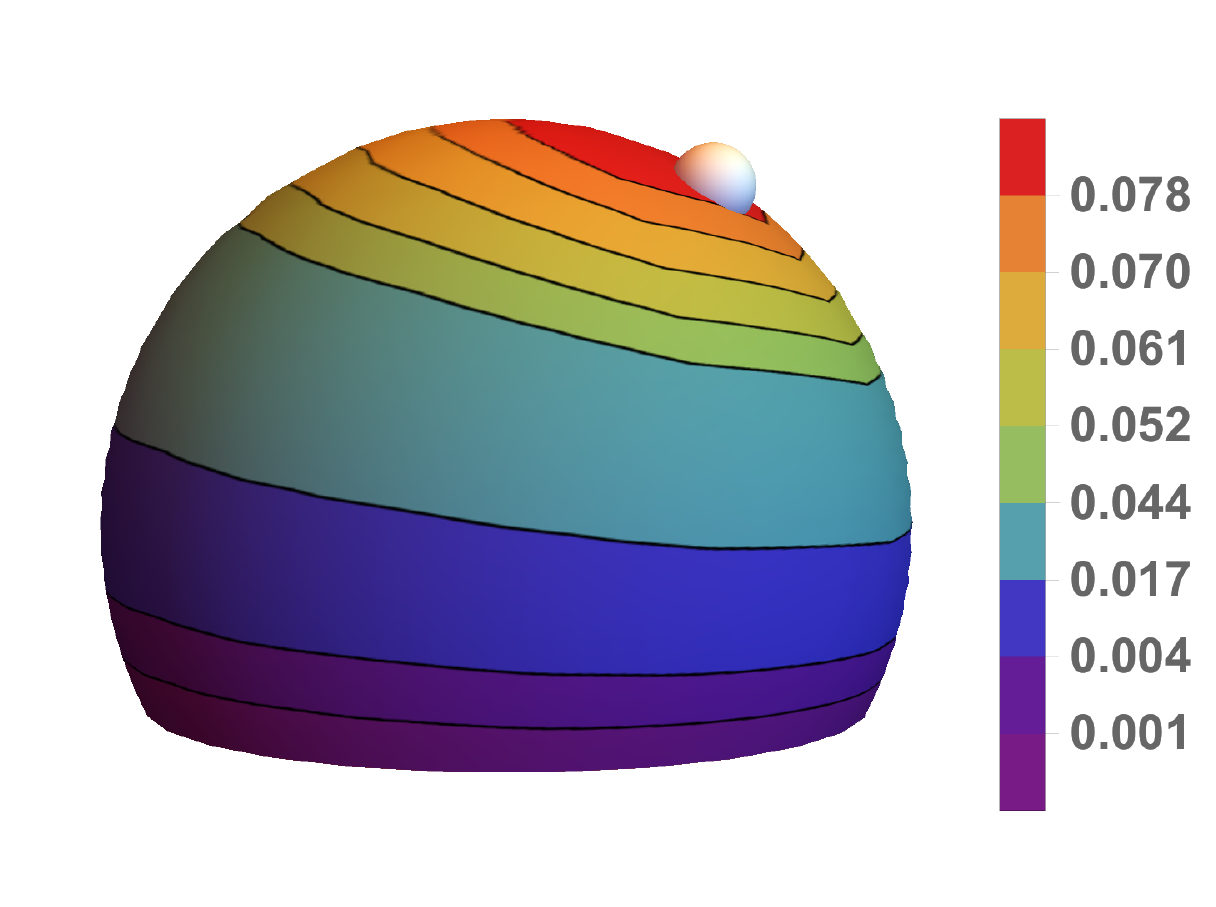}}
\caption{\footnotesize The contourplots of (a) Gaussian Curvature-Disclination interaction energy; (b) Disclination self-energy; (c) Disclination interaction energy.}
\label{energyplot}
\end{figure}

Using $G({\bf x},{\bf x}_i)$ and $\Omega({\bf x})$ in Eqs.~(\ref{G2}) and (\ref{eq:omega}) respectively, we calculate the elastic energies due to the Gaussian curvature-disclination interaction ($E_{0d}$) and disclination self-energy ($E_{self}$) as given in Eq.~(\ref{F2}).  The contourplot of the elastic free energy for $E_{0d}$ and $E_{self}$ as a function of the location of the disclination in the shell is illustrated in Fig.~\ref{energyplot}. Similar to the 1-D case shown in Fig.~4, the minimum energy due to disclination-Gaussian interaction is at the center of the shell Fig.~\ref{energyplot}(a) and related to the disclination self-energy is at the edge Fig.~\ref{energyplot}(b). While the Gaussian curvature attracts the disclination to the center of the shell, the defect self-energy pushes it towards the boundary.  The addition of these two interactions gives rise to a minimum energy, which is off the center of the shell. Figure~\ref{energyplot}(c) illustrates the contourplot of the energy of the interaction between two disclinations ($E_{dd}$).   The first disclination (the white ball in Fig.~\ref{energyplot}(c)) is placed at the minimum of $E_{0d}+E_{self}$ but the second one is located at the minimum of $E_{0d}+E_{self}+E_{dd}$.  While the curvature-disclination interaction and disclination self-energy determine the number of pentamers in the growing shell, the repulsion between the disclinations distribute them at the vertices of an icosahedron.  In this work, we calculated the interaction of Gaussian curvature with disclination and disclination self-energy in the presence of an edge for the first time.

% \begin{figure}
% \subfloat[$\sigma_g$, equation~(\ref{stressg})]{\includegraphics[width=0.5\linewidth]{stressg.pdf}}
% \subfloat[Direct $\sigma_{ds}$, equation~(\ref{stressds})]{\includegraphics[width=0.5\linewidth]{stressds.pdf}} \\
% \subfloat[Induced $\sigmba_{ds}$, equation~(\ref{stressds})]{\includegraphics[width=0.5\linewidth]{stressdirect.pdf}}
% \subfloat[Induced $\sigma_{ds}$, equation~(\ref{stressds})]{\includegraphics[width=0.5\linewidth]{stressinduced.pdf}}
% \caption{\footnotesize (a) Gaussian curvature stress contourplot; (b) Disclination stress contourplot; (c) Direct disclination stress contourplot; (d) Induced disclination stress contourplot.}
% \label{stressplot}
% \end{figure}
% \section{T=21 growth movie}

\section*{Bending and stretching Energies}
\begin{figure}
\centering
\includegraphics[width=0.7\linewidth]{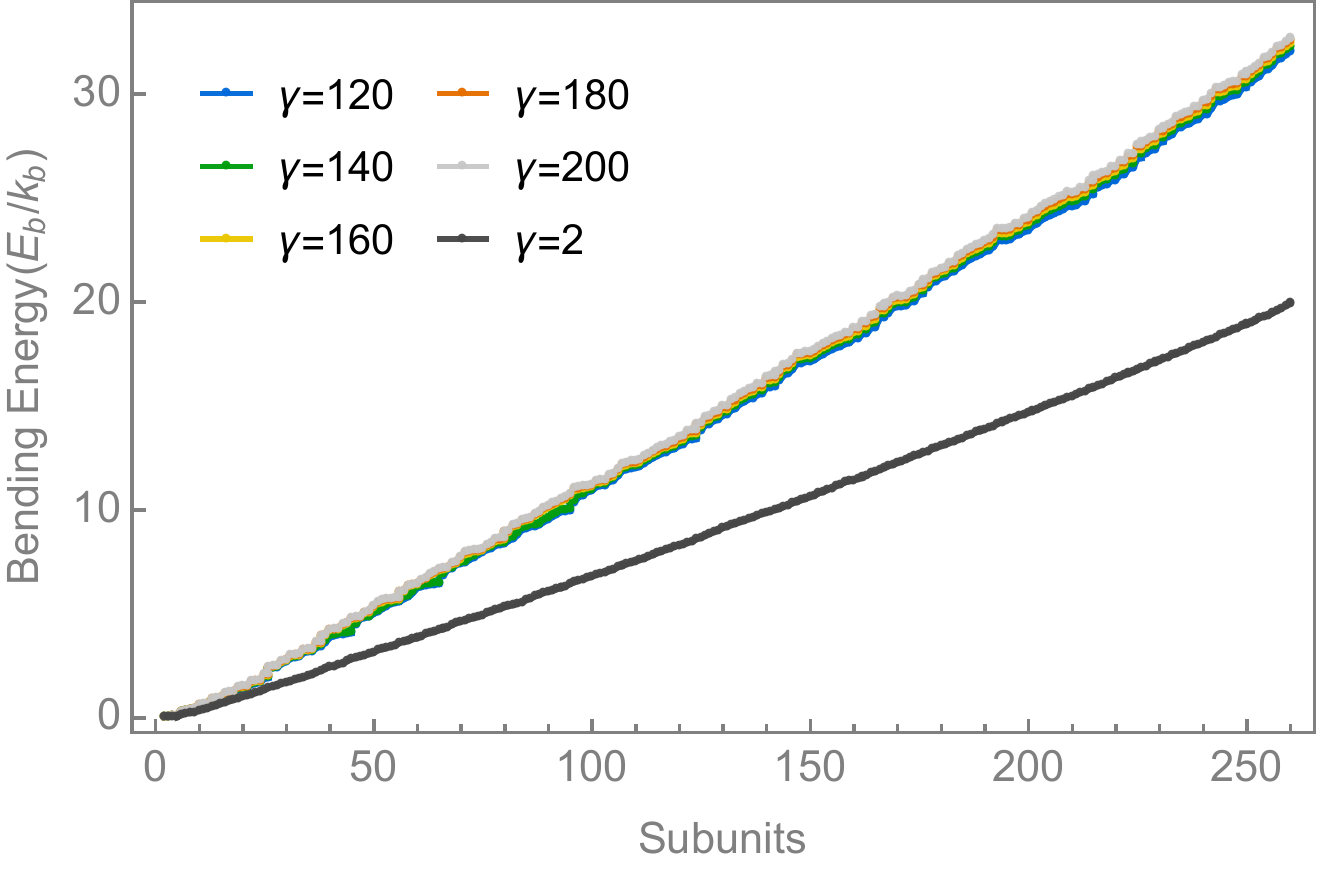}
\caption{\footnotesize The bending energy of a T=13 shell as a function of triangular subunits.}
\label{shellenergy}
\end{figure}

Figure \ref{shellenergy} illustrates the bending energy vs. the number of trimers. While there is a huge drop in the stretching energy for the formation of each pentamer(Fig.5), we see almost no difference in the bending energy between the formation of a pentamer and a hexamer as the shell grows. Note that the drops in the stretching energy other than the marked ones in Fig. 5 indicate the positions in which three or four trimers buckle out making the system ready for the assembly of a pentamer. At the later steps, when one or two trimers will be added to the growing pentamers, the stretching energy further decreases in the marked drops, revealing the formation of a complete pentamer. 

We emphasize that at large FvK ($\gamma$), the triangles remain more or less equilateral during the growth process but the shell easily bends away from the preferred curvature. In fact the magnitude of bending energy is one order of magnitude larger than stretching energy because of the deviation of dihedral angle between the proteins from the preferred one. In the simulations the bending energy was necessary to facilitate the buckling of pentamers at the right position. 

In the absence of bending rigidity, 
% if subunit-core interaction is strong, 
the pentamers 
% would be adsorbed to the shell and
became deformed to release their stress to their neighboring hexamers.
% If the subunit-core interaction is weak, the shell won’t bend to follow the core curvature and will grow as a flat sheet. 
It is interesting to note that in our system when the spontaneous radius is much smaller than the core radius, the difference between the bending energy for a pentamer and a hexamer is not considerable. We find that at high $\gamma$ the bending energy does not play any roles in the preference between pentamers and hexamers in the shell and the location of pentamers are completely controlled by the stretching energy as it is the case in the elasticity theory.

\section*{movie caption}
A step-by-step movie of the growth of a T=21 shell. The shell is built from triangular subunits. The yellow vertices belong to pentamers, blue ones to hexamers and red ones to the cap edge. The gold core mimics the preformed scaffolding layer or inner core.

\section*{Figures}

% \red{The linear fits for large and small gamma $\gamma=100.$ and $\gamma=1.0$ give expected continuum bending modulus $\kappa\approx~2.4$ and $\kappa\approx~1.5$ compared with equation (17), with correspondingly relations $\kappa=5 k_b$ and $\kappa=3 k_b$. The inconsistence comes from the increased trimer rigidity and faceted sphere, where only in small gamma the trimer subunits follow the exact curve as the inner core and obtain the same relation of two modulus as in ref...}

\bibliography{bibfile}